\documentclass[subeqn,12pt,a4paper]{article}

\usepackage{amssymb,amsmath,amsfonts,amsthm, amscd, mathrsfs,helvet, mathtools}

\usepackage{bm, stmaryrd}
\usepackage{graphicx,verbatim}
\usepackage[usenames, dvipsnames]{color}
\usepackage{psfrag}
\usepackage[all]{xy}

\usepackage{cite}

\theoremstyle{plain}

\newtheorem*{theorem*}{Theorem}

\newtheorem*{proposition*}{Proposition}

\newcommand{\tensor}[1]{{\bf \underline{#1}}}

\definecolor{brightBlue}{rgb}{0,0,1}

\definecolor{Violet}{rgb}{0.47,0,1}

 \DeclareMathOperator{\str}{Str}

\def\g{\mathfrak{g}}
\def\h{\mathfrak{h}}

\def\ha{\mbox{\small $\frac{1}{2}$}}
\def\qa{\mbox{\small $\frac{1}{4}$}}

\newcommand{\bb}[1]{\llbracket #1 \rrbracket}

\def\C{\mathcal{C}}
\def\CC{\mathbb{C}}

\def\K{\mathcal{K}}

\def\1{\tensor{1}}
\def\2{\tensor{2}}
\def\3{\tensor{3}}
\def\4{\tensor{4}}

\numberwithin{equation}{section}

\renewcommand{\L}[0]{\mathcal{L}}

\def\dss{ {\delta_{\sigma\sigma'}} }
\def\pdss{ {\partial_\sigma\delta_{\sigma\sigma'}} }

\def\beq{\begin{equation}}
\def\eeq{\end{equation}}
\def\beqz{\begin{equation*}}
\def\eeqz{\end{equation*}}
\def\bea{\begin{eqnarray}}
\def\eea{\end{eqnarray}}
\def\nn{\nonumber}
\def\et{\qquad\mbox{and}\qquad}
\def\fr{Faddeev-Reshetikhin }

\def\A{A}

\def\f{\mathfrak{f}}
\def\hf{\widehat{\mathfrak{f}}}

\def\g{\mathfrak{g}}

\def\hfs{\widehat{\mathfrak{f}}^{\sigma}}

\def\ads{$AdS_5 \times S^5$ }

\def\T{\mathcal{T}}
\def\K{\mathcal{K}}
\def\jbp{{\mathcal{J}_L}}
\def\jbm{{\mathcal{J}_R}}
\def\jb{{\mathcal{J}}}

\def\muel{{\ell}}
\def\mub{{\widetilde{\ell}}}

\begin{document}

\begin{center}
\vspace*{2em}
{\large\bf
Alleviating the non-ultralocality of the \ads superstring
}\\
\vspace{1.5em}
F. Delduc$\,{}^1$,  M. Magro$\,{}^1$, B. Vicedo$\,{}^2$

\vspace{1em}
\begingroup\itshape
{\it 1) Laboratoire de Physique, ENS Lyon
et CNRS UMR 5672, Universit\'e de Lyon,}\\
{\it 46, all\'ee d'Italie, 69364 Lyon Cedex 07, France}\\
\vspace{1em}
{\it 2) Department of Mathematics, University of York,}\\
{\it Heslington, York, YO10 5DD, United Kingdom }
\par\endgroup
\vspace{1em}
\begingroup\ttfamily
Francois.Delduc@ens-lyon.fr, Marc.Magro@ens-lyon.fr, Benoit.Vicedo@gmail.com
\par\endgroup
\vspace{1.5em}
\end{center}
\paragraph{Abstract.}

We generalize the initial steps of the Faddeev-Reshetikhin procedure to the \ads superstring theory. Specifically, we propose a modification of the Poisson bracket whose alleviated non-ultralocality enables to write down a lattice algebra for the Lax matrix. We then show that the dynamics of the Pohlmeyer reduction of the \ads superstring can be naturally reproduced with respect to this modified Poisson bracket. This work generalizes the alleviation procedure recently developed for symmetric space $\sigma$-models. It also shows that the lattice algebra recently obtained for the \ads semi-symmetric space sine-Gordon theory coincides with the one obtained by the alleviation procedure.

\section{Introduction}

It is well known since the seminal work of Bena, Polchinski and Roiban
\cite{Bena:2003wd} that classical superstring theory on \ads admits infinitely
many conserved charges. It was subsequently
shown in \cite{Magro:2008dv} that it also has infinitely many conserved
charges in involution,
thereby establishing the complete classical integrability of the theory.
But more importantly,
the result of \cite{Magro:2008dv} shows that  
 the Poisson bracket of its Lax matrix is of the
general form identified in \cite{Maillet:1985fn, Maillet:1985ek}
which is parameterized by two matrices $r$ and $s$.
The presence of the matrix  $s$ is
responsible for the non-ultralocality of this integrable field
theory  and
makes it
very problematic to define a corresponding lattice algebra. Indeed, this serious
obstacle has so far precluded the use of the
standard Quantum Inverse Scattering Method
\cite{faddtakh1979-1, Kulish:1979if, faddsklytak1980tmp1} for investigating
the quantum integrability of the \ads superstring theory.
In light of this shortcoming, the continued string of impressive developments in
this field over the past several years (see for instance the review
\cite{Beisert:2010jr})
  relied on the implicit assumption of quantum integrability in order to make
use
of the methods of factorized scattering theory \cite{Zamolodchikov:1978xm}.

However, in the case of symmetric space $\sigma$-models, we have shown in
\cite{Delduc:2012qb} how the situation may be improved by alleviating
their non-ultralocality.  This can be seen as a generalization of the first
steps of the \fr procedure \cite{Faddeev:1985qu}, developed
for the $SU(2)$ principal chiral model, to the case of symmetric space
$\sigma$-models.
Indeed, the key advantage of the alleviation procedure is that it enables to
write down a quadratic lattice algebra.
The procedure can be broken down into three parts.
The first part is achieved by purely algebraic means.
It consists in modifying the Poisson bracket of the phase space variables of the
theory in such a way that the Poisson bracket of its Lax matrix simplifies
greatly. Specifically, although the latter is still non-ultralocal,
the kernel of the new matrix $s$ is independent of spectral parameters.
Because of this, the Poisson bracket of the Lax matrix can be regularized as in
\cite{SemenovTianShansky:1995ha} and leads to a well defined lattice algebra
of the general quadratic form in \cite{Freidel:1991jx,Freidel:1991jv}.
We shall refer to such a non-ultralocality as being mild. Note that, by
construction,
the modified Poisson bracket is compatible with the original one.
The second part of the procedure concerns the degeneracy of the modified Poisson
bracket whose Casimir functions need to be determined and fixed.
Indeed, in the spirit of the \fr procedure, the purpose of the alleviation is to
reproduce the
dynamics of the $\sigma$-model with respect to the modified Poisson bracket.
However, since the latter is degenerate, only a reduction of the dynamics may be
reproduced. As shown in \cite{Delduc:2012qb}, this reduction coincides exactly
with the Pohlmeyer reduction \cite{Pohlmeyer:1975nb} of the symmetric space
$\sigma$-model. The resulting reduced dynamics is that of the symmetric
space sine-Gordon model, the Lagrangian formulation of which is given by a
gauged
Wess-Zumino-Witten model with an integrable potential \cite{Bakas:1995bm}.
The last part of the procedure consists in showing that the modified Poisson
bracket
and corresponding Hamiltonian coincide with the canonical Poisson
bracket and Hamiltonian stemming from this action.

In view of the possible generalization of the results of \cite{Delduc:2012qb}
to semi-symmetric space $\sigma$-models,
in \cite{Delduc:2012mk} we already investigated directly the canonical structure
of the semi-symmetric space sine-Gordon model obtained
by Pohlmeyer reduction of the \ads superstring
\cite{Grigoriev:2007bu,Mikhailov:2007xr}.
We have shown that the corresponding non-ultralocality is only mild and have
given the corresponding lattice algebra for the discretized Lax matrix.
The questions addressed in the present article are the following. Firstly,
does the alleviation procedure extend to the \ads superstring theory?
Secondly, is this procedure also deeply connected with the Pohlmeyer reduction?
We will find that the common answer to both questions is affirmative.

\medskip

The plan of this article is the following. In section \ref{secpbmod},
we modify the Poisson bracket of superstring theory on \ads using
a simple generalization of the technique presented in \cite{Delduc:2012qb} to
the
semi-symmetric space $F/G$, where the Lie (super)algebras respectively
associated with $F$ and $G$ are $\f = \mathfrak{psu}(2,2|4)$ and
$\g = \mathfrak{so}(4,1) \oplus \mathfrak{so}(5)$. Applying the procedure
of \cite{Delduc:2012qb} simply requires
identifying the quartet of algebraic data characterizing the integrability of
the \ads
superstring at the   Hamiltonian level. This quartet is composed of a
loop algebra, the   Hamiltonian Lax matrix of
\cite{Magro:2008dv,Vicedo:2009sn}, an $R$-matrix and an inner product.
These elements have already been identified in \cite{Vicedo:2010qd} and
therefore the modified Poisson bracket is obtained by a straightforward and
direct application of \cite{Delduc:2012qb}, namely by changing the inner
product.

Much like in the symmetric space $\sigma$-model setting, it turns out 
that most of
the constraints of the \ads superstring are Casimir functions of the 
modified
Poisson bracket. It is therefore natural to set their values to zero. 
Although some of the constraints of the \ads superstring do not correspond to 
Casimirs,
they may also be put to zero in a natural way.   Even after setting all 
of the constraints 
to zero, the
modified Poisson bracket is still degenerate.
All fields take values in $\f$ but describing the remaining Casimirs  
requires lifting one field to $G$.  
Remarkably, it turns out that  these Casimirs  correspond 
to gauge fixing conditions used in the Pohlmeyer
reduction of the \ads superstring \cite{Grigoriev:2007bu}.
We thus set their values accordingly.  Details  are given in 
section \ref{sec:casimirs}.  After summarizing the situation in 
section \ref{sec:sumpohl}, we discuss the reduced theory in section \ref{sec:
reduced}. First of all, the resulting reduced equations of
motion are exactly as in \cite{Grigoriev:2007bu} and exhibit a $H_L
\times H_R$-gauge invariance where $H_{L,R} \simeq [SU(2)]^4$.
However, they are not Hamiltonian with respect to the modified Poisson
bracket but this is remedied by partially fixing the
$H_L \times H_R$-gauge invariance to the diagonal subgroup.

We then show that these  Hamiltonian equations of motion 
 coincide with those associated
with the fermionic extension of the $G/H$ gauged WZW model with an
integrable potential as given in \cite{Grigoriev:2007bu}.
This canonical analysis is presented in section \ref{sec:gwzw}.

We conclude by some remarks. There are three appendices.  Appendix
\ref{appendixPB} contains the table of the modified Poisson bracket.  
Appendix \ref{sec:grading} recalls some important algebraic properties  
 which are 
used many times throughout this article.
 Appendix \ref{app:Ham} contains details of the derivation of the Hamiltonian.

\section{Mildly non-ultralocal Poisson bracket} \label{secpbmod}

The starting point of the procedure requires identifying the quartet of
algebraic data which encodes the integrable structure
of the \ads superstring at the Hamiltonian level.
This has been done in \cite{Vicedo:2010qd}. For completeness we briefly recall
this here and refer the reader to \cite{Delduc:2012qb} for details regarding the
present section.
The first element of this quartet is the twisted loop algebra $\hfs$
defined as follows. One starts from the Lie superalgebra
$\f = \mathfrak{psu}(2,2|4)$.  As a vector space, it admits a decomposition
into a direct sum $\oplus_{n=0}^3 \f^{(n)}$ of eigenspaces of a
$\mathbb{Z}_4$-automorphism $\sigma$ satisfying $\sigma^4 = \text{id}$.
We denote by $\g$ the Lie algebra
$\f^{(0)} = \mathfrak{so}(4,1) \oplus \mathfrak{so}(5)$ and by $G$ 
the corresponding Lie group. 
The twisted loop algebra $\hfs$ is then the
subalgebra of the
loop algebra $\hf = \f \otimes \CC(\!( \lambda )\!)$ consisting of elements
$X(\lambda) \in \hf$
which are invariant under the automorphism $\widehat{\sigma}$ of $\hf$ defined
by $\widehat{\sigma}(X)(\lambda) = \sigma[X(-i \lambda)]$. The second element
has been presented in \cite{Magro:2008dv,Vicedo:2009sn} and
is the Hamiltonian Lax matrix $\L(\lambda)$ of the theory. Its expression
in terms of the phase space variables $(A^{(i)}, \Pi^{(i)})$ reads
\begin{multline}
\L(\lambda) = A^{(0)} + \qa (\lambda^{-3} + 3\lambda) A^{(1)} + \ha(\lambda^{-2}
+ \lambda^2) A^{(2)} + \qa (3\lambda^{-1} + \lambda^3) A^{(3)} \\
+ \ha (1-\lambda^4) \Pi^{(0)} + \ha(\lambda^{-3} -\lambda) \Pi^{(1)}
+ \ha(\lambda^{-2} -\lambda^2) \Pi^{(2)} + \ha(\lambda^{-1} -\lambda^3)
\Pi^{(3)}. \label{laxmat}
 \end{multline}
The next element needed is the $R$-matrix. It is the standard one defined
by $R = \pi_{\geq 0} - \pi_{<0}$ where $\pi_{\geq 0}$ and
$\pi_{<0}$ are the projections of $\hf$ onto the subalgebras
 $\f \otimes \mathbb{C}\bb{\lambda}$ and $\f \otimes \lambda^{-1} \mathbb{C}
\bb{\lambda^{-1}}$ respectively. The last element is given by the 
twist function $\varphi(\lambda) = 4 \lambda^{-1} \phi(\lambda)$,
where the function $\phi(\lambda)$ obtained in \cite{Vicedo:2010qd} reads,
up to an irrelevant overall factor,
\beqz
\phi(\lambda) = \frac{\lambda^4}{(1-\lambda^4)^2}.
\eeqz
The twist function uniquely specifies the twisted inner product on $\hfs$,
which is defined for two elements $X$ and $Y$ of $\hfs$ by computing the residue
\beq
( X,  Y )_{\phi} = \mbox{res}_{\lambda = 0} d \lambda \frac{4}{\lambda}
 \phi(\lambda) \langle X(\lambda),  Y(\lambda) \rangle \label{pscal}
\eeq
where $\langle \cdot , \cdot \rangle$ is a non-degenerate invariant graded symmetric 
bilinear form on $\f$.

The last two elements of the quartet $(\hfs, \L, R, \varphi)$, namely the
$R$-matrix and the twist function
$\varphi$  together determine the Poisson
bracket of any two functions of the Lax matrix $\L$.
Furthermore, its non-ultralocality stems precisely from the twist function
$\varphi$ and the fact that $R$ is
not skew-symmetric with respect to \eqref{pscal} but instead satisfies
\beqz
R^\ast = - \tilde{\varphi}^{-1} \circ R \circ \tilde{\varphi} \neq - R,
\eeqz
where $\tilde{\varphi}$ denotes multiplication by $\varphi(\lambda)$.
Finally, as explained in \cite{Delduc:2012qb}, one can recover the
Poisson brackets of the fields $(A^{(i)}, \Pi^{(i)})$ appearing in the Lax matrix
\eqref{laxmat} by taking adequate functions of the
Lax matrix. The result is
\begin{subequations} \label{1109}
\begin{align}
\{ A^{(i)}_{\1}(\sigma), A^{(j)}_{\2}(\sigma')\} &=0,\\
\{ A^{(i)}_\1(\sigma), \Pi^{(j)}_{\2}(\sigma') \}&= \bigl[C^{(i \, 4-i)}_{\1\2},
A^{(i+j)}_{\2}(\sigma) \bigr]
\dss - \delta_{i+j} C^{(i \, 4-i)}_{\1\2} \pdss,\\
\{ \Pi^{(i)}_\1(\sigma), \Pi^{(j)}_\2(\sigma')  \} &= \bigl[C^{(i \,
4-i)}_{\1\2}, \Pi^{(i+j)}_\2(\sigma) \bigr] \dss,
\end{align}
\end{subequations}
where the Kronecker symbol $\delta_{i+j}$ is equal to one if
$i + j = 0 \, ({\rm mod} \; 4)$ and vanishes otherwise. Here $C_{\1\2}^{(i \, 4-i)}$
is the projection onto $\f^{(i)} \otimes \f^{(4-i)}$ of the quadratic Casimir $C_{\1\2}$.

\medskip

The alleviation procedure proposed in \cite{Delduc:2012qb} now consists in
making the following simple change in the above quartet of data
\beqz
\big( \hfs, \L,R, 4 \lambda^{-1} \phi \big) \quad \longrightarrow \quad \big(
\hfs, \L,R, 4 \lambda^{-1} \big),
\eeqz
where the factors of $4$ are introduced for later convenience.
In particular, the new quartet has the same Lax matrix as \eqref{laxmat} but a
modified Poisson bracket. The latter is still non-ultralocal as a result of the $R$-matrix
still not being skew-symmetric
\beqz
R^\ast = - \tilde{\lambda} \circ R \circ \tilde{\lambda}^{-1} \neq - R,
\eeqz
where $\tilde{\lambda}$ denotes multiplication by $\lambda$. However,
this non-ultralocality is mild in the sense that the symmetric part
$s = \ha (R + R^{\ast})$ of $R$ is a projection onto the constant part
$\f^{(0)}$ of the twisted loop algebra $\hfs$ \cite{Delduc:2012qb}.
The Poisson brackets between the various phase space fields may be
obtained from the new data along the lines of \cite{Delduc:2012qb}.
The resulting non-vanishing Poisson brackets are given in appendix
\ref{appendixPB}.

\section{Modified Poisson bracket and Pohlmeyer reduction} \label{sec:dynamics}
 
Having defined a new Poisson bracket on the phase space of the \ads
superstring, the aim of the present section will be to describe the original
dynamics with respect to it. After recalling the Hamiltonian dynamics of the
\ads superstring with respect to its original Poisson bracket \eqref{1109},
we will show that the modified Poisson bracket is degenerate so that it can
only be used to reproduce a reduction of the original dynamics. It will turn
out that the Pohlmeyer reduction is essentially forced upon us by the
specific form of the Casimirs.

\subsection{Original dynamics}

To recall the Hamiltonian dynamics of the \ads superstring we closely follow the
reference \cite{Vicedo:2009sn}. The phase space is parameterized by the fields
$(A^{(i)},\Pi^{(i)})$ and the Hamiltonian is given by a linear
combination of all the first-class constraints, namely
\beq
H = \int d\sigma \bigl[ \rho^{++} \T_{++} + \rho^{--} \T_{--}
- \str(k^{(3)} \K^{(1)}) - \str(k^{(1)} \K^{(3)})  - \str\bigl((A^{(0)} +
\muel) \C^{(0)}\bigr)\bigr], \label{hamiltonian}
\eeq
where the notation is as follows. We have defined
\begin{alignat*}{2}
\T_{++} &= T_{++} - \str\bigl(A^{(1)} \C^{(3)}\bigr), \qquad & T_{\pm\pm} &=
\str\bigl(A_\pm^{(2)} A_\pm^{(2)}\bigr),\\
\T_{--} &= T_{--} + \str\bigl(A^{(3)} \C^{(1)}\bigr), \qquad & A_\pm^{(2)}  &=
\ha\bigl( \Pi^{(2)} \mp A^{(2)} \bigr).
\end{alignat*}
The full set of constraints are
\begin{subequations} \label{constraint surface}
\begin{align}
\C^{(0)} &\equiv  \Pi^{(0)} \approx 0,\\
\C^{(1)} &\equiv \ha A^{(1)} + \Pi^{(1)}\approx 0, \label{c1}\\
\C^{(3)} &\equiv -\ha A^{(3)} + \Pi^{(3)} \approx 0, \label{c3ma}\\
T_{\pm\pm} &\approx 0. \label{vircons}
\end{align}
\end{subequations}
The constraint $\C^{(0)}$ is associated with the   
$G$-gauge invariance while \eqref{vircons} are the Virasoro constraints.
All these constraints are first-class while the other constraints $\C^{(1)}$
and $\C^{(3)}$ are partly first-class and second-class.
One can extract the following first-class constraints
\beqz
\K^{(1)} = 2i \bigl[A_-^{(2)}, \C^{(1)}]_+ \et \K^{(3)} = 2i \bigl[A_+^{(2)},
\C^{(3)}]_+,
\eeqz
which generate $\kappa$-symmetry transformations. Finally,
the arbitrary functions $\muel$, $\rho^{++}$, $\rho^{--}$, $k^{(1)}$ and
$k^{(3)}$
are Lagrange multipliers associated with the
first-class  constraints.

The equations of motion for the variables $(A^{(i)}, \Pi^{(i)})$ following from
the Hamiltonian \eqref{hamiltonian} with respect to the Poisson bracket
\eqref{1109} are, up to
terms proportional to the constraints,
\begin{subequations} \label{original eom}
\begin{align}
\partial_\tau A^{(0)} - \partial_\sigma(A^{(0)} + \muel)
- [A^{(0)} + \muel, A^{(0)}]
&= ( \rho^{++} + \rho^{--}) \bigl( \ha [A^{(2)}, \Pi^{(2)}]
+ [ A^{(1)}, A^{(3)}] \bigr)  \nn \\
&-   [A^{(1)}, Q^{(3)}]
-   [ A^{(3)}, Q^{(1)}], \label{eoma0}\\
D_\tau A^{(1)} - D_\sigma\bigl(\rho^{++} A^{(1)} +Q^{(1)}\bigr) &=
 (\rho^{++} + \rho^{--})[A^{(3)}, A_+^{(2)} ]
-   [A^{(2)}, Q^{(3)}], \label{eoma1}\\
D_\tau A_+^{(2)}   - D_\sigma\bigl( \rho^{++}
A_+^{(2)}   \bigr) &=   [A^{(1)}, Q^{(1)}], \label{eomap2}\\
D_\tau A_-^{(2)}  + D_\sigma\bigl(\rho^{--} A_-^{(2)}   \bigr) &= - [A^{(3)},
Q^{(3)}], \label{eomam2}\\
D_\tau A^{(3)} + D_\sigma\bigl(\rho^{--} A^{(3)} -Q^{(3)}\bigr) &=  (\rho^{++}
+ \rho^{--}) [A^{(1)},   A_-^{(2)} ] -   [A^{(2)}, Q^{(1)}], \label{eoma3}
\end{align}
\end{subequations}
where the covariant derivatives are defined
as
\beqz
D_\tau = \partial_\tau - [ A^{(0)} + \muel, \, ] \et
D_\sigma = \partial_\sigma - [A^{(0)}, \,].
\eeqz
Here we have also introduced the fields\footnote{The fields $Q^{(1)}$
and $Q^{(3)}$ correspond to the fields $Q_{1-}$ and $Q_{2+}$
appearing in the Lagrangian formulation \cite{Grigoriev:2007bu}.
A consequence of their definitions \eqref{defK1K3} and of the Virasoro
constraints \eqref{vircons} is that they are solutions of the algebraic
equations $[A_+^{(2)}, Q^{(1)}] =0$ and $[A_-^{(2)}, Q^{(3)}] =0$.
See also the related analysis in \cite{Vicedo:2009sn}.}
\beq
Q^{(1)} =    i[ A_+^{(2)} , k^{(1)}]_+ \et  Q^{(3)} =    i[ A_-^{(2)} ,
k^{(3)}]_+.
\label{defK1K3}
\eeq
The remaining field equations may be deduced from 
equations \eqref{original eom} by using the constraints \eqref{c1} and 
\eqref{c3ma}.
The equations of motion \eqref{original eom} are of course invariant
under the gauge transformations, which is reflected by their dependence
on arbitrary functions of $\sigma$ and $\tau$.
\medskip

\subsection{Casimirs of the modified Poisson bracket}

\label{sec:casimirs}

In order to determine whether the dynamics \eqref{original eom}
can be reproduced in terms of the modified Poisson bracket given in
appendix \ref{appendixPB}, we first need to identify the Casimirs of the latter.
Indeed, it will only be possible to reproduce a reduction of the original dynamics
where these Casimirs have been set to constants.

To begin with, $\C^{(0)}$ is an obvious Casimir of the modified
Poisson bracket. Since it corresponds to a constraint of the superstring,
the value of this Casimir is set to zero. It then follows that $\C^{(3)}$
is also a Casimir whose value we similarly set to zero.
One then finds that $A_+^{(2)}$ becomes a Casimir. This
quantity is therefore fixed to a constant by imposing
\beqz
2 A_+^{(2)} = \mu_+ T
\eeqz
where $\mu_+ \in \mathbb{R}$ is a constant and $T$ is a fixed element of
$\f^{(2)}$. But
in order for the Virasoro constraint $\str(A_+^{(2)} A_+^{(2)}) =0$ to be
satisfied,
$T$ has to be taken such that $\str T^2 =0$. We shall choose the same $T$
as in \cite{Grigoriev:2007bu}. Its definition and the fact that it induces a
$\mathbb{Z}_2$-grading of $\f$, denoted
$\f^{[0]} \oplus \f^{[1]}$, are recalled in appendix \ref{sec:grading}, along with
the definitions of some other matrices used below.

Now consider the two remaining constraints of the
\ads superstring, namely $\C^{(1)}$ and $T_{--} = \str(A_-^{(2)} A_-^{(2)})$.
Contrary to the previous constraints, these are not Casimirs of the modified
Poisson bracket. However, their only non vanishing Poisson brackets are  
\begin{subequations}
 \begin{align}
\{ \C^{(1)}_{\1}(\sigma), A^{(0)}_{\2}(\sigma') \}' &= - \ha [C_{\1\2}^{(13)},
\C^{(1)}_{\2}(\sigma) ] \dss,  \label{pbofc1} \\
 \{T_{--}(\sigma), A^{(3)}(\sigma') \}' &= - \ha [A_-^{(2)}(\sigma), \C^{(1)}(\sigma)]
\dss. \label{pbtmm}
 \end{align}
\end{subequations}
It follows from \eqref{pbofc1} that any Hamiltonian function will preserve
the constraint $\C^{(1)} =0$ with respect to the modified Poisson bracket.
 Another way to
phrase this is to note that the set of functionals on phase space which
vanish when $\C^{(1)}$ does, forms a Poisson ideal.
We may therefore restrict ourselves to the Poisson subspace defined
by $\C^{(1)} = 0$. In practice, this also means that one can take
$A^{(1)}$ as the only dynamical field belonging to $\f^{(1)}$ and identify
$\Pi^{(1)}$ with $-\ha A^{(1)}$ through equation \eqref{c1}.
Furthermore, equation \eqref{pbtmm} shows that $T_{--}$ is a Casimir
of the modified Poisson bracket on the subspace
defined by $\C^{(1)} = 0$, whose value we set to zero.
Finally, one introduces a field $g(\sigma,\tau)$ taking values in $G$
and a function $\mu_-(\sigma,\tau)$ through
\beq
2 A_-^{(2)} = \mu_- g^{-1} T g. \label{A- AdgT}
\eeq
Specifically, the polar decomposition theorem \cite{Grigoriev:2007bu,Grigoriev:2008jq} allows us
to write $2 A_-^{(2)} = g^{-1} (\mu_- T + \widetilde{\mu}_- \widetilde{T}) g$.
The vanishing of the Casimir $T_{--}$ then requires that either $\mu_- = 0$
or $\widetilde{\mu}_- = 0$. However, $\widetilde{T}$ being conjugate to $T$
by an element of $G$ (see appendix \ref{sec:grading}) equation
\eqref{A- AdgT} can be taken without loss of generality.
We are then led to consider the quantity
$\text{Str} (A_-^{(2)} A_-^{(2)} W) = -\ha \mu_-^2$. It is easily checked that,
on the subspace just defined this quantity is a Casimir function of the
modified bracket and should be put to a constant. Therefore $\mu_-$ is a
constant and the situation is thus as in \cite{Delduc:2012qb}.

However, this is not the end of the story as there exist two more
Casimirs. Indeed, consider the projection $A^{(1)[0]}$ of $A^{(1)}$ to the
subalgebra $\f^{[0]}$. We have
\beqz
\{ A^{(1)[0]}_{\1}(\sigma), A^{(1)}_{\2}(\sigma')\}' = - \ha [C_{\1\2}^{(13)[00]},
A_{+\2}^{(2)}]\dss = - \qa \mu_+ [C_{\1\2}^{(13)[00]}, T_{\2}]\dss = 0,
\eeqz
as any element of $\f^{[0]}$ commutes with $T$ (see appendix 
\ref{sec:grading}), and where $C_{\1\2}^{(13)[00]}$ denotes the projection
onto $\f^{(1)[0]} \otimes \f^{(3)[0]}$ of $C_{\1\2}^{(13)}$.
All the other Poisson brackets with $A^{(1)[0]}$ either vanish as well or are
proportional to $\C^{(1)}$, which in practice has the same consequence.
In other words $A^{(1)[0]}$ is a Casimir. This is a nice result as it
corresponds to one of the gauge fixing conditions for the $\kappa$-symmetry
considered in \cite{Grigoriev:2007bu}. The other condition will also be
encountered shortly. In order to describe it explicitly we first need
to lift the Poisson brackets of $A_-^{(2)}$ to the field $g$. 
This lifting is done as follows. The only non-vanishing Poisson bracket of
$A_-^{(2)}$ is
\beqz
\{ A_{-\1}^{(2)}(\sigma), A^{(0)}_{\2}(\sigma') \}' = - \ha [C_{\1\2}^{(22)},
A^{(2)}_{-\2}] \dss.
\eeqz
This may be lifted using \eqref{A- AdgT} to
a Poisson bracket for $g$ which reads
\beqz
\{ g_{\1}(\sigma), A^{(0)}_{\2}(\sigma') \}' = - \ha g_{\1}(\sigma)
C_{\1\2}^{(00)} \dss,
\eeqz
with all the other Poisson brackets of $g$ vanishing. 
Next, the only non-vanishing Poisson brackets of $A^{(3)}$ are
\begin{align*}
\{ A_{\1}^{(3)}(\sigma), A^{(0)}_{\2}(\sigma') \}' &=
- \ha [C_{\1\2}^{(31)}, A^{(3)}_{\2}(\sigma)] \dss,\\
\{ A_{\1}^{(3)}(\sigma), A^{(3)}_{\2}(\sigma') \}' &=
- \ha [C_{\1\2}^{(31)}, A^{(2)}_{-\2}(\sigma)] \dss.
\end{align*}
Considering the combination $g A^{(3)} g^{-1}$, a short computation leads to
\begin{subequations} \label{pbofpsi3}
\begin{align}
\{(g A^{(3)} g^{-1})_{\1}(\sigma), A^{(0)}_{\2}(\sigma') \}' &= 0,\\
\{ (g A^{(3)} g^{-1})_{\1}(\sigma), (g A^{(3)} g^{-1})_{\2}(\sigma') \}' &=
- \ha [C_{\1\2}^{(31)}, (g_{\2}A^{(2)}_{-\2}g_{\2}^{-1})(\sigma)] \dss
= - \qa \mu_{-} [C_{\1\2}^{(31)}, T_{\2}] \dss.
 \end{align}
\end{subequations}
As in the case of $A^{(1)[0]}$ above
this shows that $(g A^{(3)} g^{-1})^{[0]}$ is a Casimir, which exactly
corresponds to the other gauge fixing condition for $\kappa$-symmetry
considered in \cite{Grigoriev:2007bu}.

\subsection{Pohlmeyer reduction} \label{sec:sumpohl}

Let us summarize the situation so far. We have shown that the modified
Poisson bracket given in appendix \ref{appendixPB} can be consistently
restricted to the constraint surface of the \ads superstring defined by
\eqref{constraint surface}. But this restriction is still degenerate and
the form of its Casimirs naturally led us to impose the following
further conditions
\begin{subequations} \label{Pohlred}
\begin{equation}
2 A_+^{(2)} = \mu_+ T \et 2 A_-^{(2)} = \mu_- g^{-1} T g \label{reda2}
\end{equation}
along with
\begin{equation}
A^{(1)[0]} = 0 \et (gA^{(3)} g^{-1})^{[0]} =0. \label{reda13}
\end{equation}
\end{subequations}
These are exactly the gauge fixing conditions imposed in the Pohlmeyer
reduction of the \ads superstring \cite{Grigoriev:2007bu}. In other words,
the modified Poisson bracket naturally restricts to the reduced phase space
of the Pohlmeyer reduction of the \ads superstring.
It is easy to check that the gauge fixing conditions \eqref{Pohlred} are
preserved
under the dynamics if 
\begin{equation}
\rho^{++} = 1, \qquad \rho^{--} = 1, \qquad Q^{(1)} = 0, \qquad Q^{(3)} = 0,
\qquad \muel(\sigma, \tau) \in \h. \label{GTcond}
\end{equation}
These equations are also partial gauge fixing conditions imposed in
\cite{Grigoriev:2007bu},
to which we refer the reader for further detail.

The remaining degrees of freedom are $g$, $A^{(0)}$, $A^{(1)[1]}$
and $(g A^{(3)} g^{-1})^{[1]}$ and their non-vanishing Poisson brackets read 
\begin{subequations} \label{finalpb}
 \begin{align}
 \{ g_{\1}(\sigma), A^{(0)}_{\2}(\sigma') \}' &= - \ha g_{\1}(\sigma)
C_{\1\2}^{(00)} \dss,\\
\{ A^{(0)}_{\1}(\sigma), A^{(0)}_{\2}(\sigma') \}' &=
- \ha [C_{\1\2}^{(00)}, A^{(0)}_{\2}(\sigma) ] \dss + \ha C_{\1\2}^{(00)} \pdss,
\\
 \{ A^{(1)[1]}_{\1}(\sigma), A^{(1)[1]}_{\2}(\sigma')\}' &= - \qa \mu_+
[C_{\1\2}^{(13)},
T_{\2}]\dss,\\
 \{ (g A^{(3)} g^{-1})^{[1]}_{\1}(\sigma), (g A^{(3)}
g^{-1})^{[1]}_{\2}(\sigma') \}' &=
- \qa \mu_{-} [C_{\1\2}^{(31)}, T_{\2}] \dss.
 \end{align}
\end{subequations}

\subsection{Reduced equations of motion} \label{sec: reduced}

Next, we implement the reduction conditions \eqref{Pohlred} together with
\eqref{GTcond} on the equations of motion \eqref{original eom} in turn.
For the equation \eqref{eoma0} of $A^{(0)}$ we find
\beq
\partial_-A^{(0)} - \partial_\sigma  \muel -[ \muel, A^{(0)}] =
\ha \mu_+ \mu_- [g^{-1} T g, T] + 2 [A^{(1)}, A^{(3)}], \label{eqrzero}
\eeq
where $\partial_{\pm} = \partial_{\tau} \pm \partial_{\sigma}$.
Equation \eqref{eomam2} can be lifted to an equation of motion for $g$,
exactly as in the bosonic case, to give
\beq
A^{(0)} = \ha \bigl( - g^{-1} \partial_+ g -  \muel + g^{-1}
 \mub  g\bigr), \label{eqra0}
\eeq
where the arbitrary function $\mub$ takes values in $\h$. On the odd graded part
of $\f$,
the equation \eqref{eoma1} for $A^{(1)}$ yields
\beq
\partial_- A^{(1)} = [ \muel, A^{(1)}] +  \mu_+ [A^{(3)}, T]. \label{eqrreda1}
\eeq
As for the equation of motion \eqref{eoma3} of $A^{(3)}$, using
\eqref{eqra0} it may be rewritten as
\beq
\partial_+ (g A^{(3)} g^{-1}) =
[\mub , gA^{(3)} g^{-1}] +  \mu_- [gA^{(1)} g^{-1}, T]. \label{eqredlast}
\eeq
Note that the projections of equations \eqref{eqrreda1} and \eqref{eqredlast}
to $\f^{[0]}$ are both trivial, therefore we shall implicitly assume their
restrictions to $\f^{[1]}$ from now on.

The equations of motion \eqref{eqrzero}-\eqref{eqredlast} admit
right and left gauge invariances. The right invariance
corresponds to those $\g$-gauge transformations that
preserve the reduction conditions. They act as  
\begin{subequations} \label{gtr}
\begin{gather}
\delta A^{(0)} = \partial_\sigma \alpha_R +[\alpha_R,A^{(0)}], \quad
\delta A^{(1)} = [\alpha_R,A^{(1)}], \quad
\delta A^{(3)} =[\alpha_R,A^{(3)}], \\
\delta g = -g \alpha_R, \quad
\delta  \muel =  \partial_- \alpha_R +[\alpha_R,  \muel], 
\end{gather}
\end{subequations}
where $\alpha_R(\sigma,\tau) \in \mathfrak{h}_R$. There is also
a left invariance which appears as a result of the lifting to $G$.
It acts only on the fields $g$ and $\mub$ as
\beq
 \delta g  = \alpha_L g \et
\delta  \mub  = \partial_+ \alpha_L + [\alpha_L,  \mub ], \label{gtl}
\eeq
with $\alpha_L(\sigma,\tau) \in \mathfrak{h}_L$.

To obtain equations of motion
that are Hamiltonian, one needs to partially gauge fix this
$H_L \times H_R$-gauge invariance to the diagonal subgroup.
To do this, we introduce
\beq
J = \partial_\sigma g g^{-1} + g A^{(0)} g^{-1}. \label{defofJ}
\eeq
A short computation shows that $J$ satisfies the
following equation of motion
\beqz
\partial_+ J  = \partial_\sigma  \mub  + [\mub  , J]
+ \ha \mu_+ \mu_- [T, g T g^{-1}] + 2 g [A^{(1)} , A^{(3)} ] g^{-1},
\eeqz
and has the following Poisson brackets
\begin{align*}
\{ J_{\1}(\sigma), g_{\2}(\sigma') \}' &= \ha C_{\1\2}^{(00)} g_{\2}(\sigma)
\dss,\\
\{ J_{\1}(\sigma), A_{\2}^{(0)}(\sigma') \}' &= 0,\\
\{ J_{\1}(\sigma), A_{\2}^{(1)}(\sigma') \}'&= 0,\\
\{ J_{\1}(\sigma), (gA^{(3)}g^{-1})_{\2}(\sigma') \}'&= 0.
 \end{align*}
With the help of this field $J$ we may now write the generator of the gauge
transformations \eqref{gtr} and \eqref{gtl} explicitly as follows
\beqz
2 \int d\sigma \str \left[ \alpha_L \left(J
+\frac{1}{\mu_-} \big[ gA^{(3)} g^{-1}, [T, gA^{(3)} g^{-1}] \big] \right)
- \left( A^{(0)} - \frac{1}{\mu_+} \big[ A^{(1)}, [T, A^{(1)}] \big] \right) \alpha_R
\right].
\eeqz
We therefore fix the part of the gauge invariance with parameters
related through $\alpha_L = - \alpha_R$ by
imposing the partial gauge fixing condition
\beq
J^{[0]} + \frac{1}{\mu_-} \big[ g A^{(3)} g^{-1}, [T, g A^{(3)} g^{-1}] \big] =
A^{(0)[0]} - \frac{1}{\mu_+} \big[ A^{(1)}, [T, A^{(1)}] \big]. \label{gfc}
\eeq
The residual gauge transformations that preserve this condition are
the diagonal transformations for which $\alpha_L = \alpha_R$.
Moreover, condition \eqref{gfc} is preserved by the dynamics
\eqref{eqrzero}-\eqref{eqredlast} provided  the arbitrary 
functions $\muel$ and $\mub$ are restricted as
\beq
\muel -  \mub = -A^{(0)[0]} - J^{[0]} + \frac{1}{\mu_+} \big[A^{(1)}, [T, A^{(1)}] \big]
- \frac{1}{\mu_-} \big[ g A^{(3)} g^{-1}, [T, g A^{(3)} g^{-1}] \big]. \label{cgfc}
\eeq
Equations \eqref{gfc} and \eqref{cgfc} can be rearranged into
the equivalent set of equations
\begin{subequations} \label{ell tell}
\begin{align}
\muel &= \ha( \muel +  \mub)  - A^{(0)[0]} +
\frac{1}{\mu_+} \big[ A^{(1)}, [T, A^{(1)}] \big], \label{101}\\
\mub &= \ha(\muel +  \mub)+ J^{[0]} +
\frac{1}{\mu_-} \big[ g A^{(3)} g^{-1}, [T, g A^{(3)} g^{-1}] \big]. \label{100}
\end{align}
\end{subequations}
In other words, after imposing the condition \eqref{gfc},
the equations of motion no longer depend on the pair
of arbitrary functions $\muel$ and $\mub$ but only on their sum $\muel + \mub$.
This is a reflection of the fact that the equations of motion are
invariant only under the diagonal gauge transformations.

To implement the partial gauge fixing conditions \eqref{gfc} at the level
of the equations of motion we simply need to substitute the relations
\eqref{ell tell} for $\muel$ and $\mub$.
The equations of motion \eqref{eqrreda1} and \eqref{eqredlast}
for the fermionic fields respectively yield
\begin{subequations} \label{f1}
\begin{align}
\partial_- A^{(1)} &= - \mu_+ [T, A^{(3)}]
+\left[ \ha ( \muel +  \mub) - A^{(0)[0]} +
\frac{1}{\mu_+} \big[ A^{(1)}, [T, A^{(1)}] \big], A^{(1)}\right],\\
\partial_+(g A^{(3)} g^{-1}) &= - \mu_-[T, g A^{(1)} g^{-1}]  \nn \\
&\qquad \qquad + \left[ \ha ( \muel +  \mub) + J^{[0]}
+ \frac{1}{\mu_-} \big[ g A^{(3)} g^{-1}, [T, g A^{(3)} g^{-1}] \big],
g A^{(3)} g^{-1} \right].
 \end{align}
\end{subequations}
For the equation of $g$ we first combine equations \eqref{eqra0}
and \eqref{defofJ} to get
\beqz
\partial_\tau g g^{-1} + J + g(A^{(0)} +  \muel)g^{-1}  =  \mub.
\eeqz
Then substituting both expressions \eqref{100} and \eqref{101} into this
equation we end up with
\begin{multline}
 \partial_\tau g = -g A^{(0)[1]} - J^{[1]} g -
g \left(\ha( \muel +  \mub) + \frac{1}{\mu_+} \big[ A^{(1)},[T, A^{(1)}] \big] \right) \\
+\left( \ha( \muel +  \mub) + \frac{1}{\mu_-}
\big[ g A^{(3)} g^{-1}, [T, g A^{(3)}g^{-1}] \big] \right) g \label{eqgfinal}
\end{multline}
Finally, the equation of motion \eqref{eqrzero} can be rewritten as
\begin{multline}
\partial_\tau A^{(0)} =
\partial_\sigma A^{(0)[1]} + \partial_\sigma  \left(
\ha( \muel +  \mub) + \frac{1}{\mu_+} \big[ A^{(1)},[T, A^{(1)}] \big] \right)
+ \ha \mu_+ \mu_- [g^{-1}Tg,T]\\
+ \left[\ha( \muel +  \mub) - A^{(0)[0]} +\frac{1}{\mu_+} \big[ A^{(1)},[T,
A^{(1)}] \big],
A^{(0)}\right] + 2[A^{(1)}, A^{(3)}]\label{f2}
\end{multline}
where again we have made use of \eqref{101}.

\section{Link with semi-symmetric space sine-Gordon theory} \label{sec:gwzw}

The goal of this section is to establish that the Poisson brackets
\eqref{finalpb} and the constraint \eqref{gfc} coincide with
the result of the canonical analysis of the
\ads semi-symmetric space sine-Gordon theory, defined as a fermionic extension
of the
$G/H$ gauged WZW with a potential term \cite{Grigoriev:2007bu}.
In order to make the identification complete, we also indicate the
corresponding Hamiltonian which
generates the equations of motion \eqref{f1},
\eqref{eqgfinal} and \eqref{f2}. 

We shall perform the canonical analysis of the action defined in
\cite{Grigoriev:2007bu}
which reads
\begin{align}
\mathcal{S} &=  \ha \int d\tau d\sigma \str(g^{-1} \partial_+ g g^{-1}
\partial_- g) + \mbox{\small $\frac{1}{3}$} \int d\tau d\sigma d\xi
\epsilon^{\alpha \beta \gamma}
\str( g^{-1} \partial_\alpha g g^{-1} \partial_\beta g g^{-1} \partial_\gamma g)
\nn\\
& - \int d\tau d\sigma \str(B_+ \partial_- g g^{-1} - B_- g^{-1}
\partial_+ g   + g^{-1} B_+ g B_- - B_+ B_-)  \nn \\
& + \ha \int d\tau d\sigma
\str(\psi^{(3)} [T, D_+ \psi^{(3)}] + \psi^{(1)} [T, D_- \psi^{(1)}]) \nn\\
&  + \int d\tau d\sigma \bigl(
\mu^2 \str(g^{-1}TgT) + \mu \str(g^{-1} \psi^{(3)} g \psi^{(1)}) \bigr),
\label{WZWaction}
\end{align}
where the notation is as follows. Firstly, we take $\epsilon^{\tau \sigma \xi} =
1$.
The fields $g$, $\psi^{(1)}$ and $\psi^{(3)}$ respectively take values in
$G$, $\f^{(1)[1]}$ and $\f^{(3)[1]}$, while $B_\pm = B_0 \pm B_1$ are
gauge fields taking values in $\h$. Finally, the covariant derivatives are
defined by $D_\pm = \partial_\pm - [B_\pm, ]$.
We recall the start of the canonical analysis from the results of
\cite{Delduc:2012mk}. 
The phase space is parametrized by the fields
$(g,\jbp,\psi^{(1)},\psi^{(3)})$ where $\jbp$ takes values in $\g$, and the
non-vanishing Poisson brackets are
\begin{subequations} \label{pbofgwzw}
\begin{align}
\{ g_{\1}(\sigma) , \jb_{L\2}(\sigma')  \}' &= g_{\1} C_{\1\2}^{(00)} \dss, \\
\{ \jb_{L\1}(\sigma), \jb_{L\2}(\sigma') \}' &= [C_{\1\2}^{(00)}, \jb_{L\2}]
\dss
+ 2 C_{\1\2}^{(00)}
\pdss,\\
 \{ \psi^{(1)}_{\1}(\sigma), \psi^{(1)}_{\2}(\sigma') \}' &= \bigl[ T_{\2},
C_{\1\2}^{(13)}
\bigr] \dss,\\
\{ \psi^{(3)}_{\1}(\sigma), \psi^{(3)}_{\2}(\sigma') \}' &= \bigl[ T_{\2},
C_{\1\2}^{(31)}
\bigr] \dss,
\end{align}
\end{subequations}
together with the gauge fields $(B_0,B_1)$ and their conjuguate 
momenta\footnote{Their Poisson bracket is canonical, \emph{i.e.}
$\{B_{0\1}(\sigma), P_{0\2}(\sigma')\}'= C_{\1\2}^{(00)[00]} \dss$ 
and similarly for $B_1$ and $P_1$.} $(P_0,P_1$). 
There are four constraints, 
\begin{subequations} \label{fourc}
\begin{gather}
\chi_1 =  P_0 \et \chi_2 = P_1,\\
\chi_3 = \jb_R^{[0]} + 2 B_1 - \ha \big[\psi^{(3)}, [T, \psi^{(3)}]\big], 
\label{defchi3}\\
\chi_4 = \jb_L^{[0]} + 2 B_1 + \ha \big[\psi^{(1)}, [T, \psi^{(1)}]\big]
\label{defchi4}
\end{gather}
\end{subequations}
where we have defined
\beqz
\jbm = -2 \partial_\sigma g g^{-1} + g \jbp g^{-1}.
\eeqz
To achieve the comparison with the previous section, we first put
strongly to zero the set of second-class constraints $\chi_2$ and $\chi_3$. 
In addition, we fix the gauge invariance generated by the 
first-class constraint $\chi_1$ by imposing the condition $B_0 = 0$.
All this is done by introducing the corresponding Dirac bracket and
by explicitly eliminating the variables $(B_1,P_1)$ and $(B_0,P_0)$.
In particular, the elimination of $B_1$ is realized using the definition
\eqref{defchi3} of $\chi_3$ to make the replacement
\beq
B_1 \to -\ha \jbm^{[0]} + \qa \big[\psi^{(3)}, [T, \psi^{(3)}]\big].
\label{replacea1}
\eeq
The result of this procedure is a straightforward
generalization to the case at hand of the result obtained in
 \cite{Bowcock:1988xr}. The Dirac brackets for the remaining fields
$(g,\jbp,\psi^{(1)},\psi^{(3)})$ are the same as
their Poisson brackets. We are left with the single constraint $\chi_4$ which 
according to the rule \eqref{replacea1} becomes
\beq
\chi_4 = \jb_L^{[0]} - \jb_R^{[0]}
+ \ha \big[\psi^{(1)}, [T, \psi^{(1)}]\big]
+ \ha \big[\psi^{(3)}, [T, \psi^{(3)}]\big]. \label{schi4}
\eeq
The corresponding Hamiltonian is computed in appendix \ref{app:Ham} and reads
\begin{multline}
H' = \int \!\! d\sigma \str\Bigl[  \,\,  \qa  \bigl( \jbp^{[1]} \jbp^{[1]}
+ \jbm^{[1]} \jbm^{[1]}\bigr) - \ha \psi^{(3)} [T, \partial_{\sigma} \psi^{(3)}]
+ \ha\psi^{(1)} [T, \partial_{\sigma} \psi^{(1)}] - \mu^2 g^{-1}TgT \\
- \mu g^{-1} \psi^{(3)} g \psi^{(1)}
- \mbox{\small $\frac{1}{16}$} \big[\psi^{(3)}, [T, \psi^{(3)}]\big]
\big[\psi^{(3)}, [T, \psi^{(3)}]\big]
- \mbox{\small $\frac{1}{16}$} \big[\psi^{(1)}, [T, \psi^{(1)}]\big]
\big[\psi^{(1)}, [T, \psi^{(1)}]\big]\\
- \qa \jbp^{[0]} \big[\psi^{(1)}, [T, \psi^{(1)}]\big]
+ \qa \jbm^{[0]} \big[\psi^{(3)}, [T, \psi^{(3)}]\big]
+ \lambda  \chi_4 \Bigr] \label{veryfinH}
\end{multline}
where $\lambda$ is a Lagrange multiplier. 

In summary, the phase space of the \ads semi-symmetric space sine-Gordon theory
may be parametrized by the fields $(g,\jbp,\psi^{(1)},\psi^{(3)})$ with Poisson
brackets
given in \eqref{pbofgwzw} and subject to the first-class constraint
\eqref{schi4}.
So we are now in a position to give the sought dictionary
between section \ref{sec:dynamics} and the present section. As suggested by the
notation,
the field $g$ and the constant matrix $T$ are the same in both sections, whereas
the
remaining fields and parameters are related by  
\begin{alignat}{2} \label{thedico}
\jbp &= - 2 A^{(0)}, &\qquad \jbm &= - 2 J, \nn \\
\psi^{(1)} &= \frac{2}{\sqrt{\mu_+}} A^{(1)[1]}, &\qquad
\psi^{(3)} &= \frac{2}{\sqrt{\mu_-}} (g A^{(3)} g^{-1})^{[1]}, \\
\mu &= - \sqrt{\mu_+\mu_-}, &\qquad
\lambda &= -\ha( \muel + \mub). \nn
\end{alignat}
One can check that there is perfect agreement, firstly between
the Poisson brackets \eqref{pbofgwzw} and \eqref{finalpb},
secondly between the constraints \eqref{schi4}
and \eqref{gfc}, and lastly between the
equations of motion generated by the Hamiltonian \eqref{veryfinH}
and the equations of motion \eqref{f1}, \eqref{eqgfinal} and \eqref{f2}.

\section{Conclusion}

Let us start by answering the questions which motivated this work as mentioned
in the introduction. We have shown that the alleviation procedure,
as developed in \cite{Delduc:2012qb} for symmetric space $\sigma$-models, extends
smoothly to the case of the \ads superstring. Moreover, we have found that in
this context as well the procedure is tightly linked with Pohlmeyer reduction. 

\medskip

An important point we wish to stress concerns the rigidity of the alleviation
procedure. Indeed, at every stage of the procedure there is essentially no freedom.
To begin with, the introduction of the modified Poisson bracket is guided by the
requirement that its non-ultralocality be only mild. This
places severe restrictions on the choice of inner product entering the definition
of the Poisson bracket. Subsequently, the degeneracy of the modified
Poisson bracket and the specific form of its Casimirs basically compel us to
restrict attention to the phase space of the Pohlmeyer reduction of the
\ads superstring. The complete procedure therefore leads us very naturally
from the \ads superstring theory to the associated semi-symmetric space
sine-Gordon theory.

\medskip

By comparison with our previous work \cite{Delduc:2012qb} where 
we were not considering a string theory, let us briefly recall that
in the context of the \ads superstring theory, Pohlmeyer reduction
corresponds to a reduction of gauge degrees of freedom.
The reduction therefore still describes the dynamics
of all the physical degrees of freedom of the original \ads superstring.
Of course, in the bosonic setting the same interpretation
holds if, say, for the $\sigma$-model on $S^n$ we consider instead a
string theory on $\mathbb{R} \times S^n$ (see for instance
\cite{Grigoriev:2007bu, Miramontes:2008wt}). 

\medskip

One could of course take the canonical structure of the \ads superstring
and consider its own restriction to the reduced degrees of freedom.
In the context of the \ads superstring, this problem has been addressed
first in \cite{Mikhailov:2006uc} and later in more details in 
\cite{Schmidtt:2010bi,Schmidtt:2011nr}. It turns out that the 
induced Poisson structure is non-local. This is in stark contrast
with the restriction of the modified Poisson bracket to the reduced
degrees of freedom as presented in this article. Indeed, the latter is
perfectly local but more importantly
it has the property that the corresponding Poisson bracket of the
Lax matrix is mildly non-ultralocal.

\medskip

Evidently, the equivalence between the original \ads superstring and the
theory with the modified Poisson bracket describing the Pohlmeyer 
reduction
of the \ads superstring is only classical at this stage. Whether or not
this equivalence persists at the quantum level is likely to be a rather 
delicate issue.
Indeed, the corresponding statement for the $SU(2)$ principal 
chiral model in
\cite{Faddeev:1985qu} requires a subtle change of vacuum from 
the reference state
of the Bethe ansatz to the physical ground state given by the Dirac 
sea of Bethe roots.
To further this program, the next  challenge would be to find the 
quantization of the quadratic lattice algebra of the Lax matrix as described
in \cite{Delduc:2012mk}.

\paragraph{Acknowledgements}

We thank A. Le Diffon for comments on the draft.
B.V. is supported by UK EPSRC grant EP/H000054/1.

\appendix

\section{Modified Poisson bracket}
\label{appendixPB}
We reproduce below the modified Poisson bracket, which is mildly
non-ultralocal. The only Poisson bracket, which involves a 
derivative of the Dirac $\delta$-function is   
\beqz
\{ A^{(0)}_{ \1}(\sigma), A^{(0)}_{ \2}(\sigma') \}' = -
\ha [C^{(00)}_{\1\2}, A^{(0)}_{\2} + \ha \C^{(0)}_{\2}] \delta_{\sigma \sigma'} + \ha
C^{(00)}_{\1\2} \pdss.  
\eeqz
The complete list of all the other non-vanishing Poisson brackets 
is
\beqz
\begin{array}{ll}
\{ A^{(0)}_{ \1}(\sigma), A^{(1)}_{ \2}(\sigma') \}' = - \qa
[C^{(00)}_{\1\2}, \C^{(1)}_{\2}] \delta_{\sigma \sigma'}, &
\{ A^{(0)}_{ \1}(\sigma), A^{(2)}_{ \2}(\sigma') \}' = - \ha
[C^{(00)}_{\1\2}, \A^{(2)}_{-\2}] \delta_{\sigma \sigma'}, \cr
\{ A^{(0)}_{ \1}(\sigma), A^{(3)}_{ \2}(\sigma') \}' = - \ha
[C^{(00)}_{\1\2}, A^{(3)}_{\2} + \ha \C^{(3)}_{\2}] \delta_{\sigma \sigma'}, &
\{ A^{(1)}_{ \1}(\sigma), A^{(1)}_{ \2}(\sigma') \}' = - \ha
[C^{(13)}_{\1\2}, \A^{(2)}_{+\2}] \delta_{\sigma \sigma'},\cr
\{ A^{(1)}_{ \1}(\sigma), A^{(2)}_{ \2}(\sigma') \}' = - \qa
[C^{(13)}_{\1\2}, \C^{(3)}_{\2}] \delta_{\sigma \sigma'},&
\{ A^{(1)}_{ \1}(\sigma), A^{(3)}_{ \2}(\sigma') \}' = - \qa
[C^{(13)}_{\1\2}, \C^{(0)}_{\2}] \delta_{\sigma \sigma'},\cr
\{ A^{(2)}_{ \1}(\sigma), A^{(2)}_{ \2}(\sigma') \}' = - \qa
[C^{(22)}_{\1\2}, \C^{(0)}_{\2}] \delta_{\sigma \sigma'},&
\{ A^{(2)}_{ \1}(\sigma), A^{(3)}_{ \2}(\sigma') \}' = - \qa
[C^{(22)}_{\1\2}, \C^{(1)}_{\2}] \delta_{\sigma \sigma'},\cr
\{ A^{(3)}_{ \1}(\sigma), A^{(3)}_{ \2}(\sigma') \}' = - \ha
[C^{(31)}_{\1\2}, \A^{(2)}_{-\2}] \delta_{\sigma \sigma'},&
\{ A^{(0)}_{ \1}(\sigma), \Pi^{(1)}_{\2}(\sigma') \}' = - \mbox{\small
$\frac{3}{8}$}
[C^{(00)}_{\1\2}, \C^{(1)}_{\2}] \delta_{\sigma \sigma'},\cr
\{ A^{(0)}_{ \1}(\sigma), \Pi^{(2)}_{\2}(\sigma') \}' = - \ha
[C^{(00)}_{\1\2}, \A^{(2)}_{-\2}] \delta_{\sigma \sigma'},&
\{ A^{(0)}_{ \1}(\sigma), \Pi^{(3)}_{\2}(\sigma') \}' = - \qa
[C^{(00)}_{\1\2}, A^{(3)}_{\2} + \ha \C^{(3)}_{\2}] \delta_{\sigma \sigma'},\cr
\{ A^{(1)}_{ \1}(\sigma), \Pi^{(1)}_{\2}(\sigma') \}' = \qa
[C^{(13)}_{\1\2}, \A^{(2)}_{+\2}] \delta_{\sigma \sigma'},&
\{ A^{(1)}_{ \1}(\sigma), \Pi^{(2)}_{\2}(\sigma') \}' = \qa
[C^{(13)}_{\1\2}, \C^{(3)}_{\2}] \delta_{\sigma \sigma'},\cr
\{ A^{(1)}_{ \1}(\sigma), \Pi^{(3)}_{\2}(\sigma') \}' = \mbox{\small
$\frac{3}{8}$}
[C^{(13)}_{\1\2}, \C^{(0)}_{\2}] \delta_{\sigma \sigma'},&
\{ A^{(2)}_{ \1}(\sigma), \Pi^{(1)}_{\2}(\sigma') \}' = \mbox{\small
$\frac{1}{8}$}
[C^{(22)}_{\1\2}, \C^{(3)}_{\2}] \delta_{\sigma \sigma'},\cr
\{ A^{(2)}_{ \1}(\sigma), \Pi^{(2)}_{\2}(\sigma') \}' = \qa
[C^{(22)}_{\1\2}, \C^{(0)}_{\2}] \delta_{\sigma \sigma'},&
\{ A^{(2)}_{ \1}(\sigma), \Pi^{(3)}_{\2}(\sigma') \}' = - \mbox{\small
$\frac{1}{8}$}
[C^{(22)}_{\1\2}, \C^{(1)}_{\2}] \delta_{\sigma \sigma'},\cr
\{ A^{(3)}_{ \1}(\sigma), \Pi^{(1)}_{\2}(\sigma') \}' = \mbox{\small
$\frac{1}{8}$}
[C^{(31)}_{\1\2}, \C^{(0)}_{\2}] \delta_{\sigma \sigma'},&
\{ A^{(3)}_{ \1}(\sigma), \Pi^{(2)}_{\2}(\sigma') \}' = - \qa
[C^{(31)}_{\1\2}, \C^{(1)}_{\2}] \delta_{\sigma \sigma'},\cr
\{ A^{(3)}_{ \1}(\sigma), \Pi^{(3)}_{\2}(\sigma') \}' = - \qa
[C^{(31)}_{\1\2}, \A^{(2)}_{-\2}] \delta_{\sigma \sigma'},&
\{ \Pi^{(1)}_{\1}(\sigma), \Pi^{(1)}_{\2}(\sigma') \}' = - \mbox{\small
$\frac{1}{8}$}
[C^{(13)}_{\1\2}, \A_{+\2}^{(2)}] \delta_{\sigma \sigma'},\cr
\{ \Pi^{(1)}_{\1}(\sigma), \Pi^{(2)}_{\2}(\sigma') \}' = - \mbox{\small
$\frac{1}{8}$}
[C^{(13)}_{\1\2}, \C^{(3)}_{\2}] \delta_{\sigma \sigma'},&
\{ \Pi^{(1)}_{\1}(\sigma), \Pi^{(3)}_{\2}(\sigma') \}' = - \mbox{\small
$\frac{3}{16}$}
[C^{(13)}_{\1\2}, \C^{(0)}_{\2}] \delta_{\sigma \sigma'},\cr
\{ \Pi^{(2)}_{\1}(\sigma), \Pi^{(2)}_{\2}(\sigma') \}' = - \qa
[C^{(22)}_{\1\2}, \C^{(0)}_{\2}] \delta_{\sigma \sigma'},&
\{ \Pi^{(2)}_{\1}(\sigma), \Pi^{(3)}_{\2}(\sigma') \}' = - \mbox{\small
$\frac{1}{8}$}
[C^{(22)}_{\1\2}, \C^{(1)}_{\2}] \delta_{\sigma \sigma'}, \cr
\{ \Pi^{(3)}_{\1}(\sigma), \Pi^{(3)}_{\2}(\sigma') \}' =- \mbox{\small
$\frac{1}{8}$}
[C^{(31)}_{\1\2}, \A^{(2)}_{-\2}] \, \delta_{\sigma \sigma'}.&
\end{array}
\eeqz

\section{Additional $\mathbb{Z}_2$-grading} \label{sec:grading}

Besides the $\mathbb{Z}_4$-grading of $\f$ introduced in section \ref{secpbmod},
throughout the article we make extensive use of an additional
$\mathbb{Z}_2$-grading of $\f$ \cite{Grigoriev:2007bu}. 
We list here its definition and main properties. 

We follow the conventions of \cite{Grigoriev:2007bu} 
with regards to the Lie superalgebra $\mathfrak{psu}(2,2|4)$. Defining the
matrix
\begin{equation} \label{defT}
T = \mbox{\small $\frac{i}{2}$} \text{diag}(1,1,-1,-1,1,1,-1,-1),
\end{equation}
it can be used to define a $\mathbb{Z}_2$-grading $\f = \f^{[0]}
\oplus \f^{[1]}$
by setting
\beq
\f^{[0]} = \{ M \in \f \,|\, [T, M] = 0 \}, \qquad
\f^{[1]} = \{ M \in \f \,|\, [T, M]_+ = 0 \}. \label{f0f1def}
\eeq
The projectors onto the respective spaces in \eqref{f0f1def} are given by
\beq
M^{[0]}  = - [T,[T,M]_+]_+ \et M^{[1]} = - [T,[T,M]]. \label{pro01}
\eeq
Note that $\f^{[0]} = \text{Ker}(\text{ad}\, T)$ and an alternative
characterization
of $\f^{[1]}$ is given by $\f^{[1]} = \text{Im}(\text{ad}\, T)$. This leads at
once to $\str( \mathfrak{f}^{[0]} \mathfrak{f}^{[1]} ) = 0$.

The subspace $\f^{(2)[0]}$ is two dimensional, and defining the matrix
\begin{equation*}
W = \text{diag}(1,1,1,1,-1,-1,-1,-1),
\end{equation*}
it is spanned by $T$ and $\widetilde{T} = W T$. The matrix $\widetilde{T}$
is conjugate to $T$ by an element of $G$ \cite{Grigoriev:2008jq}.

\section{Derivation of the Hamiltonian} \label{app:Ham}

In this appendix we derive the Hamiltonian \eqref{veryfinH} governing the
dynamics
of the \ads semi-symmetric space sine-Gordon theory, after eliminating the
constraints $\chi_2$, $\chi_3$ explicitly and gauge fixing the invariance
generated by $\chi_1$.

The Hamiltonian obtained from the action \eqref{WZWaction} by Legendre
transform reads
\begin{align}
 H'&= \int \!\! d\sigma  \str\Bigl[ \,\,  \qa \bigl( \jbp^2 + \jbm^2\bigr)
- \ha \psi^{(3)} [T, \partial_{\sigma} \psi^{(3)}] + \ha \psi^{(1)} [T,
  \partial_{\sigma} \psi^{(1)}] \nn\\
& - \mu^2 g^{-1}TgT
- \mu g^{-1} \psi^{(3)} g \psi^{(1)}+ \jbm (B_0 +B_1) - \jbp  (B_0 -B_1)  + 2
 B_1^2 \nn\\
&+ \ha \psi^{(3)} \bigl[T, [ (B_0 + B_1), \psi^{(3)}]\bigr] + \ha \psi^{(1)}
\bigl[T, [ (B_0 -
B_1), \psi^{(1)}]\bigr] \Bigr].
\end{align}
One can use the definitions \eqref{defchi3} and \eqref{defchi4}
of the constraints $\chi_3$ and $\chi_4$ to rewrite this as
\begin{multline}
 H'= \int \!\! d\sigma  \str\Bigl[\,\,  \qa \bigl( \jbp^2 + \jbm^2\bigr)
- \ha \psi^{(3)} [T, \partial_{\sigma} \psi^{(3)}] + \ha \psi^{(1)} [T,
  \partial_{\sigma} \psi^{(1)}]  - \mu^2 g^{-1}TgT
- \mu g^{-1} \psi^{(3)} g \psi^{(1)} \\
+ B_0 (\chi_3 - \chi_4)
+ B_1 (\chi_3 + \chi_4 - 2 B_1) \Bigr]. \label{lastlineh}
\end{multline}
We may add to the Hamiltonian density a term proportional to the square
of any constraint since this has no effect on the dynamics along
the constraint surface. Adding $- \qa \str(\chi^2_4)$, the last two terms in
\eqref{lastlineh} may then be rewritten as
\begin{equation*}
B_0 (\chi_3 - \chi_4) + B_1 (\chi_3 + \chi_4 - 2 B_1) - \qa \chi^2_4 =
(B_0 + B_1) \chi_3 - B_0 \chi_4 - \big( \ha \chi_4 - B_1 \big)^2 - B_1^2.
\end{equation*}

As explained in section \ref{sec:gwzw}, we may impose the constraint
$\chi_3 = 0$ strongly by introducing a Dirac bracket for the constraints
$\chi_2$ and $\chi_3$. Using the explicit expression \eqref{defchi4} for
$\chi_4$ we have
$\ha \chi_4 - B_1 = \ha \jbp^{[0]} + \qa \big[\psi^{(1)}, [T, \psi^{(1)}]\big]$.
We should then also replace $B_1$ by the expression
in \eqref{replacea1}.
Putting all of
this together we obtain the Hamiltonian governing the
dynamics of the remaining fields
\begin{multline}
H'= \int \!\! d\sigma \str\Bigl[  \,\,  \qa  \bigl( \jbp^{[1]} \jbp^{[1]}
+ \jbm^{[1]} \jbm^{[1]}\bigr) - \ha \psi^{(3)} [T, \partial_{\sigma} \psi^{(3)}]
+ \ha\psi^{(1)} [T, \partial_{\sigma} \psi^{(1)}] - \mu^2  g^{-1}TgT\\
- \mu  g^{-1} \psi^{(3)} g \psi^{(1)}
- \mbox{\small $\frac{1}{16}$} \big[\psi^{(3)}, [T, \psi^{(3)}]\big]
\big[\psi^{(3)}, [T, \psi^{(3)}]\big]
- \mbox{\small $\frac{1}{16}$} \big[\psi^{(1)}, [T, \psi^{(1)}]\big]
\big[\psi^{(1)}, [T, \psi^{(1)}]\big]\\
- \qa  \jbp^{[0]} \big[\psi^{(1)}, [T, \psi^{(1)}]\big]
+ \qa \jbm^{[0]} \big[\psi^{(3)}, [T, \psi^{(3)}]\big] - B_0  \chi_4
\Bigr].  
\end{multline}
One can check 
that it preserves the constraint $\chi_4$.
At this point there remains two gauge invariances generated by the
first-class constraints $\chi_1$ and $\chi_4$. We therefore add to the
Hamiltonian 
density the linear combination $\str (v_0 \chi_1  + \lambda \chi_4)$ 
where $v_0$ and $\lambda$ are Lagrange multipliers. We fix the 
invariance generated by $\chi_1$ by imposing the condition $B_0=0$. Preserving
this constraint requires $v_0=0$ and we arrive at the Hamiltonian
\eqref{veryfinH}.

\providecommand{\href}[2]{#2}\begingroup\raggedright\endgroup

\end{document}